\documentclass{article}
\usepackage{spconf}
\usepackage{amsmath,amssymb,amsfonts,amsthm}
\usepackage{algorithm, algpseudocode}
\usepackage{graphicx}
\usepackage{enumitem, color}
\usepackage[numbers, sort&compress]{natbib}
\usepackage[hidelinks]{hyperref}
\input{mysymbol.sty}

\title{Bayesian topology inference on partially known networks from input-output pairs}
\name{Martín Sevilla and Santiago Segarra}
\address{
Electrical and Computer Engineering, Rice University, USA
}
\begin{document}
\ninept
\maketitle
\begin{abstract}
We propose a sampling algorithm to perform system identification from a set of input-output graph signal pairs.
The dynamics of the systems we study are given by a partially known adjacency matrix and a generic parametric graph filter of unknown parameters.
The methodology we employ is built upon the principles of annealed Langevin diffusion.
This enables us to draw samples from the posterior distribution instead of following the classical approach of point estimation using maximum likelihood.
We investigate how to harness the prior information inherent in a dataset of graphs of different sizes through the utilization of graph neural networks.
We demonstrate, via numerical experiments involving both real-world and synthetic networks, that integrating prior knowledge into the estimation process enhances estimation performance.
\let\thefootnote\relax\footnotetext{Research was sponsored by the Army Research Office and was accomplished under Grant Number W911NF-17-S-0002.
We also gratefully acknowledge financial support by the Fulbright U.S. Student Program, which is sponsored by the U.S. Department of State and the U.S.--Argentina Fulbright Commission.
The views and conclusions contained in this document are those of the authors and should not be interpreted as representing the official policies, either expressed or implied, of the Army Research Office, the U.S. Army, the Fulbright Program, the U.S.--Argentina Fulbright Commission or the U.S. Government. 
The U.S. Government is authorized to reproduce and distribute reprints for Government purposes notwithstanding any copyright notation herein.
Emails: \url{msevilla@rice.edu}, \url{segarra@rice.edu}}
\end{abstract}
\section{Introduction}
\label{sec:intro}
Graphs are widely used modeling tools for complex networks across numerous disciplines and can serve as the domain of data supported on their set of nodes. 
In essence, this allows us to conceptualize signals as residing on a graph~\cite{ortega2018graph}. 
Although in various practical scenarios the underlying graph supporting the given data is known, in many other instances, one has to estimate it. 
This leads to the problem of \emph{network topology inference}, where the objective is to learn the graph structure based on observations at the nodes. 
A multitude of diverse approaches exist today to address this challenge within the realm of graph signal processing and beyond~\cite{Mateos_2019, Dong_2019}.

Our focus in this paper is on performing system identification based on a set of input-output observations.
Specifically, we assume that the system is given by a graph filter~\cite{ortega2018graph} of unknown parameters applied to a partially known adjacency matrix.
The motivation behind this setup is twofold.
First, linking an input and an output graph signal in this manner is a generic way to understand a process in a graph that can incorporate any kind of network diffusion, including heat diffusion, consensus, and opinion dynamics~\cite{Mateos_2019}.
Hence, we aim to devise an algorithm that is applicable to \emph{any type} of graph filters and networks.
Second, we allow our algorithm to deal with partially known graphs so that edges that are \emph{known} to either exist or be absent can be incorporated as prior knowledge.
This is useful in many practical cases, including gene expression data~\cite{weighted_glasso_1}, neurotoxicology tests~\cite{grzebyk2004identification}, social networks~\cite{Wu852798} and brain network models~\cite{SIMPSON2015310}.
Furthermore, we tackle this network inference problem from a novel Bayesian standpoint, allowing us to incorporate prior information on the underlying graph structure.
It is usual to count on a dataset of graphs whose distribution is known to be the same, regardless of it not presenting a closed-form expression to work with.
Here we model the prior using a graph neural network (GNN) trained on the graphs dataset, and leverage Langevin dynamics, a stochastic diffusion process, to render sampling from a difficult posterior distribution possible.

\vspace{1mm}
\noindent {\bf Relation to prior work.}
Plenty of work has been carried out recently in order to tackle the task of graph learning from nodal observations.
In some works, the network structure is learned from graph signals by constraining the graph to belong to a certain set of graphs (e.g., trees~\cite{lu2018learning} or product graphs~\cite{einizade2023learning}).
Similarly, other works tackle the problem by assuming that the graph filters are of a specific kind (e.g., they assume consensus dynamics~\cite{Zhu_2020} or heat diffusion filters~\cite{thanou2016learning}).
In~\cite{egilmez2018graph, kalofolias2016learn, 7178669,7990048,shafipour2019identifying,madeline}, the functional form of the graph filter is generic, but the signals are assumed to be smooth on the underlying graph or generated by passing Gaussian noise through the generic filter.
Additionally, all of these works deal with output signals and assume that knowledge regarding the input is not available, which makes the problem they solve different from the one we study in this work.
Graph learning from input-output pairs has been studied in~\cite{8450838}, but they restrict the mapping between the signals to be given by an autoregressive moving-average graph filter.
Additionally, to the best of our knowledge, none of the previous work on network topology inference has dealt with the incorporation of structural prior information about the graph, other than sparsity or structural information that can be directly encoded as spectral constraints~\cite{rey2023enhanced, kumar2019structured}.
In~\cite{ji2023graph}, a distribution of graphs is considered to design graph signal processing tools applicable to several graphs simultaneously. 
However, this significantly differs from our goal of recovering a single graph shaped by a prior distribution of graphs.

\vspace{1mm}
\noindent {\bf Contributions.}
The contributions of this paper are threefold:
\begin{enumerate}[wide, labelindent=0pt]
    \item We introduce a novel algorithm to estimate a system's dynamics given by a network process on a partially known graph.
    Our algorithm centers on sampling from a posterior distribution, in contrast to the conventional approach of seeking the maximum \textit{a posteriori} or the maximum likelihood estimators.
    \item We implement this estimator leveraging annealed Langevin dynamics. 
    This approach enables us to incorporate diverse prior distributions learned from graph datasets, even if they have different numbers of nodes.
    Additionally, the method allows us to employ any differentiable graph filter to model input-output relationships.
    \item Our numerical experiments showcase that the incorporation of prior knowledge surpasses approaches that solely enforce sparsity or rely exclusively on the likelihood of the observed signals.
\end{enumerate}

\section{Problem formulation}
Let $\ccalG$ be a graph of $N$ nodes represented by its adjacency matrix $\bbA$.
In the context of this work, our focus is on unweighted (i.e. $A_{ij} \in \{ 0, 1\}$) and undirected (i.e. $A_{ij} = A_{ji}$) graphs with no self-loops (i.e., $A_{ii} = 0$).
We consider $\ccalG$ to have a partially known set of edges, so that some entries $A_{ij}$ are \emph{known} to be either $0$ or $1$, and we aim to estimate the rest.
To differentiate between the known entries of $\bbA$ and the entries we intend to estimate, we define two sets of indices $\ccalO$ and $\ccalU$ such that
\begin{equation}
    \ccalO = \left\{(i,j) : A_{ij} \ \text{is observed} \land i < j\right\}
\end{equation}
and
\begin{equation}
    \ccalU = \left\{(i,j) : A_{ij} \ \text{is unknown} \land i < j\right\} .
\end{equation}
Throughout this work, we refer to the known and unknown fractions of the adjacency matrix as $\bbA^\ccalO$ and $\bbA^\ccalU$, respectively.

Let $\bby \in \reals^N$ be a measured output graph signal that corresponds to a known input $\bbx \in \reals^N$ such that
\begin{equation} \label{eq:system}
    \bby = h_{\bbtheta} (\bbA) \bbx + \bbvarepsilon,
\end{equation}
where $\bbvarepsilon \sim \ccalN (\pmb{0}, \sigma_{\varepsilon}^2 \bbI)$ is measurement noise, and $h_{\bbtheta}(\cdot)$ is a known graph filter with unknown parameters $\bbtheta$ (i.e., we only know its functional form).
We assume that $K$ independent input-output signal pairs are available, and that the relationship between them is given by~\eqref{eq:system}.
For convenience, we arrange the input and output graph signals into the matrices $\bbX =\begin{bmatrix} \bbx_1 \ \ldots \ \bbx_K \end{bmatrix} \in \reals^{N \times K}$ and $\bbY =\begin{bmatrix} \bby_1 \ \ldots \ \bby_K \end{bmatrix}  \in \reals^{N \times K}$, respectively.

Additionally, we want to incorporate some prior information into our estimator.
Usually, sparsity is the only prior knowledge that is used when carrying out topology inference.
We want, however, to be able to use \emph{any} other type of prior information available.
In order to do this, we assume that a set of adjacency matrices $\ccalA$ is available, where each matrix corresponds to a graph whose prior distribution $p(\bbA)$ matches that of the graph we aim to estimate.
In other words, in our setup, prior information is given by a set of graphs that are assumed to belong to the same distribution as $\ccalG$.
It is important to note that our approach does, therefore, not rely on having a closed-form expression for $p(\bbA)$ in any manner. 
Instead, we work with a collection of adjacency matrices that might not even share the same dimensions, since graphs from the same distribution can have different numbers of nodes.
In this setting, our problem is defined as follows:
\begin{problem}\label{prob:main}
    Given the $K$ graph signal pairs $\bbX$ and $\bbY$, a partially known adjacency matrix $\bbA^\ccalO$, and prior knowledge given by a set of adjacency matrices $\ccalA$, find an estimate of $\bbA^\ccalU$ and $\bbtheta$ knowing that the dynamics are given by~\eqref{eq:system}.
\end{problem}
A natural way of solving Problem~\ref{prob:main} is to compute the maximum \textit{a posteriori} (MAP).
Such estimator must obviously equal $\bbA^\ccalO$ for all positions $(i,j) \in \ccalO$.
The MAP is computed by solving the following optimization problem:
\begin{alignat}{2}
    \label{eq:map}
    \hbA_\MAP, \hbtheta_{\MAP}
    = 
    &\argmax_{\bbA, \bbtheta} & \ & p(\bbA, \bbtheta \!\mid\! \bbX, \bbY) \nonumber \\
    =&\argmax_{\bbA, \bbtheta} & \ & p\left(\bbX, \bbY \mid \bbA, \bbtheta \right) p\left(\bbA, \bbtheta\right) \\
    &\text{subject to} & & A_{ij} = A^\ccalO_{ij}, \,\,\, \forall (i,j) \in \ccalO. \nonumber
\end{alignat}
In general, $\bbA$ and $\bbtheta$ are independent and hence $p(\bbA, \bbtheta) = p(\bbA)p(\bbtheta)$.
Nonetheless, there is an important obstacle that prevents us from solving~\eqref{eq:map}: a closed-form expression for $p(\bbA)$ is not available, since we just know $\ccalA$, a set of samples from $p(\bbA)$.
Furthermore, as we aim to be able to solve Problem~\ref{prob:main} for a \emph{generic} family of graph filters (i.e., we do not look for an approach that works with a specific functional form for $h(\cdot)$), the expression for the likelihood $p(\bbX, \bbY \!\mid\! \bbA, \bbtheta)$ alone may not even be tractable from an optimization standpoint.

The main question that arises is how to exploit the predictive power given by $\ccalA$, and how to be able to deal with \emph{any} filter $h(\cdot)$.
In this work, we propose to change the scope of the problem.
Rather than finding the values that maximize the posterior probability (i.e., solving~\eqref{eq:map}), our approach consists of \emph{drawing a sample} from the posterior distribution instead.
While the obtained sample may not always coincide with the exact MAP estimate, it provides a practical and effective solution. 
In fact, the sampled $\bbA^\ccalU$ is chosen with a high probability of being a representative point from the posterior distribution, offering a computationally feasible approximation to~\eqref{eq:map}.

\section{Proposed method}
\subsection{Annealed Langevin sampling} \label{subsec:langevin}
We sample from the posterior distribution $p(\bbA, \bbtheta \!\mid\! \bbX, \bbY)$ by leveraging annealed Langevin dynamics, a stochastic diffusion process.
The Langevin dynamics algorithm serves as a Markov chain Monte Carlo (MCMC) technique, enabling the generation of samples from distributions that are challenging to directly sample from~\cite{MCMCbook, Roberts1996ExponentialCO}.
For a target distribution $p(\bbw)$, consider the recursion
\begin{equation} \label{eq:langevin_dt}
	\bbw_{t+1} = \bbw_t + \epsilon \nabla_{\bbw_t}\log p(\bbw_t) + \sqrt{2\epsilon\tau}\, \bbz_t,
\end{equation}
where $t$ is an iteration index, $\epsilon$ is a step size, $\tau$ is the \textit{temperature} parameter, and $\bbz_t \sim \ccalN(0, \bbI)$. 
This discrete process is known as the unadjusted Langevin algorithm (ULA).

The dynamics governed by~\eqref{eq:langevin_dt} have an intuitive explanation. 
During each iteration, $\bbw_t$ is inclined to move in the direction of the gradient of the logarithm of the target density (known as the \textit{score function}). 
However, the movement is also influenced by white noise $\bbz_t$, which prevents the point from getting trapped in local maxima. 
Given certain regularity conditions, as $\epsilon \to 0$ and $t \to \infty$, $\bbw_t$ converges to a sample drawn from the distribution $p(\bbw)^{1/\tau}$~\cite{wellinglang}.
Selecting $\tau = 1$ gives rise to samples drawn from the target distribution $p(\bbw)$.
Choosing $\tau \neq 1$ can be particularly useful for multimodal distributions~\cite{pavliotis2016stochastic}.

It is worth pointing out that the \emph{only} requirement to sample from $p(\bbw)$ using this method is knowing the score function $\nabla \log p(\bbw)$.
We revisit this useful advantage of the ULA in Section~\ref{subsec:gnn}.
However, it is important to acknowledge that, in our scenario, we are aiming to sample a \emph{discrete} random vector (specifically, a vectorized unweighted adjacency matrix).
Consequently, the gradient of the target log-density is undefined. 
To approach this challenge, the idea is to use a noisy (continuous) variant of these random variables instead of the original discrete vector. 
Such an approach gives rise to the concept of \emph{annealed} Langevin dynamics~\cite{annealed_langevin, kawar2021snips, nicoz_1, nicoz_2}.

To simplify the notation of what follows, we work with the half-vectorization of the adjacency matrix.
In a slight abuse of notation, we use $\bba = \mathrm{vech}(\bbA)$ and $\bbA$ interchangeably.
Let $\{\sigma_l\}_{l=1}^{L}$ be a sequence of \emph{noise levels} such that $\sigma_1 > \sigma_2 > \cdots > \sigma_L > 0$. 
Then, for each noise level, we define a noisy version of $\bba$ at the iteration $t$ of Langevin,
\begin{equation}\label{eq:pert_symbs}
    \tba_t = \bba + \bbn_{l(t)},
\end{equation}
where $\bbn_l \sim \ccalN(\pmb{0}, \sigma_l^2 \bbI)$. 
In this context, the iterative process involving the application of annealed Langevin dynamics to our problem is described by the following recursion:
\begin{equation} \label{eq:annealed_langevin}
	\tba_{t+1} = \tba_t + \alpha_t \nabla_{\tba_t}\log p(\tba_t, \bbtheta | \bbX, \bbY) + \sqrt{2\alpha_t \tau}\, \bbz_t,
\end{equation}
where $\alpha_t = \epsilon \cdot \sigma_{l(t)}^2 / \sigma_L^2$.
Notice that the noise present in $\tba_t$ decreases with $t$, since we stay at the same noise level for a fixed number of iterations $T$ before switching to the next (smaller) one (see Algorithm~\ref{alg:annealed_langevin}).

The introduction of the annealed version of the dynamics was initially aimed at enhancing the overall performance of the algorithm~\cite{annealed_langevin}.
Nonetheless, within our specific problem, this version also presents the benefit of rendering the target distribution differentiable.
When appropriately selecting the sequence $\{\sigma_l\}_{l=1}^{L}$ and the step size $\epsilon$, after a sufficient number of iterations, the continuous sample $\tba_t$ becomes arbitrarily close to an actual sample drawn from the discrete distribution $p(\bba)$.

Since each $\sigma_l$ is a predefined parameter, the sequence must be designed so that the first noise levels are high enough to explore the whole space, while the last ones are rather small, in order to approximate the true discrete distribution given by a set of scaled Dirac deltas at all possible values of $\bba$.
Additionally, setting the parameter temperature at $\tau < 1$ allows us to highlight more the peaks of the original discrete distribution in the landscape of the annealed, continuous one.

In order to compute the gradient needed to run the recursion in~\eqref{eq:annealed_langevin}, we express the posterior distribution as
\begin{align} \label{eq:joint_posterior}
    p(\tbA, \bbtheta \mid \bbY, \bbX)
    &=
    \frac{p(\bbY \mid \tbA, \bbtheta, \bbX)p(\tbA)}{p(\bbY)},
\end{align}
where we have assumed independence between $\tbA$ and $\bbtheta$, and we have also assigned a flat improper prior to $p(\bbtheta)$.
Taking the gradient of~\eqref{eq:joint_posterior} with respect to $\tbA$ we get
\begin{equation} \label{eq:posterior_score}
    \scalemath{0.97}
    {\nabla \log p(\tbA, \bbtheta \mid \bbY, \bbX)
    =
    \nabla \log p(\bbY \mid \tbA, \bbtheta, \bbX) + \nabla \log p(\tbA)}.
\end{equation}

In order to use~\eqref{eq:posterior_score} to sample $\tbA$ using~\eqref{eq:annealed_langevin}, we need to be able to compute the two terms involved: the score of the likelihood and the score of the annealed prior.
The former is straightforward, as $(\bby_k \mid \tbA, \bbtheta, \bbx_k) \sim \ccalN(h_{\bbtheta} (\tbA) \bbx_k, \sigma_{\varepsilon}^2 \bbI)$.
Since each measurement is independent of the rest, $p(\bbY \mid \tbA, \bbtheta, \bbX)$ is just the product of $K$ multivariate normal densities.
This leads to the log-likelihood being
\begin{equation} \label{eq:score_likelihood}
    \log p(\bbY \mid \tbA, \bbtheta, \bbX) 
    = - \frac{1}{2 \sigma_\varepsilon^2} \sum_{k=1}^{K} \norm{\bby_k - h_{\bbtheta}(\tbA)\bbx_k}_2^2.
\end{equation}
If $h(\cdot)$ is differentiable with respect to $\tba$ and $\bbtheta$, the expression in~\eqref{eq:score_likelihood} is differentiable with respect to both variables as well, regardless of the functional form of the graph filter. 
We compute the derivatives via automatic differentiation~\cite{autodiff}.

Conversely, computing the second term in~\eqref{eq:posterior_score} poses a notably more challenging task. 
Specifically, $p(\tbA)$ characterizes the density of a vector governed by a discrete distribution, yet affected by white Gaussian noise in some of its components. 
Furthermore, in our situation, we do not even count on an expression for $p(\bbA)$; rather, we rely solely on the collection of graphs $\ccalA$.
Fortunately, it is possible to \emph{approximate} the annealed prior score function by devising an estimator such that $\bbs\left(\tba, \sigma\right) \approx \nabla \log p\left(\tba\right)$ using neural networks~\cite{annealed_langevin}.

\subsection{Learned annealed scores} \label{subsec:gnn}
Let us recall that all we need is to estimate $\nabla \log p(\tbA)$ so that we can sample from~\eqref{eq:joint_posterior} by using the diffusion in~\eqref{eq:annealed_langevin}.
Let $\bbs_{\bbxi}(\tba, \sigma)$ be the output of the GNN we aim to train, with $\bbxi$ being its learnable parameters.
Ideally, the output for a given $\tba$ (with the corresponding noise level $\sigma_l$ for the current iteration) should closely approximate the actual score function $\nabla \log p(\tba)$.
The loss function designed to learn $\bbxi$ needs to be constructed in a way that simultaneously minimizes the mean squared error (MSE) across all noise levels.
To achieve this, we define the loss function
\begin{equation}
    \ccalJ\left(\bbxi\mid\{\sigma_l\}_{l=1}^L\right)
    =
    \frac{1}{2L}\sum_{l=1}^L \sigma_l^2 
    \E{\ccalD\left(\tba\mid\bbxi, \sigma_l\right)},
    \label{eq:loss_cond}
\end{equation}
where
\begin{equation} \label{eq:dist_cond}
\begin{split}
    \ccalD\left(\tba\mid\bbxi, \sigma_l\right)
    &=
    \norm{\bbg_{\bbxi} (\tba, \sigma_l) - \nabla \log p(\tba\mid\bba)}_2^2 \\
    &=
    \norm{\bbg_{\bbxi} (\tba, \sigma_l) - (\bba - \tba)/\sigma_l^{2}}_2^2.
\end{split}
\end{equation}
Using the score function of $p(\tba \mid \bba)$ instead of $p(\tba)$ in~\eqref{eq:dist_cond} returns the same minimizer as using the latter~\cite{scorematching_autoencoders}. 
Hence, the output of a GNN trained with~\eqref{eq:loss_cond} correctly estimates $\nabla \log p(\tba)$.
It is important to highlight that the term $(\bba - \tba)/\sigma_l^2$ is entirely known during training: $\bba$ constitutes the ground-truth graph examples in the dataset, while both $\tba$ and $\sigma_l$ correspond to the inputs of the network.

In this study, we use the EDP-GNN architecture~\cite{edpgnn}.
This network employs a permutation equivariant approach to effectively model the desired score function, and has been tailored to specifically tackle score-matching tasks on graphs.

\subsection{Final algorithm}
So far, we have focused on estimating the adjacency matrix $\bbA$ but not the filter parameters $\bbtheta$.
Because we do not assume any prior information on $\bbtheta$, we just aim to maximize the likelihood of the observed signals given the parameters.
Namely, from the standpoint of optimizing $\bbtheta$, maximizing the posterior distribution in~\eqref{eq:joint_posterior} is equivalent to minimizing the loss function
\begin{equation} \label{eq:theta_loss}
    \ccalL(\bbtheta) = - \log p(\bbY \mid \tbA, \bbtheta, \bbX),
\end{equation}
whose expression was given in~\eqref{eq:score_likelihood}.

\begin{figure*}
    \centering
    \begin{minipage}{0.49\textwidth}
        \centering
        \includegraphics[width=\textwidth]{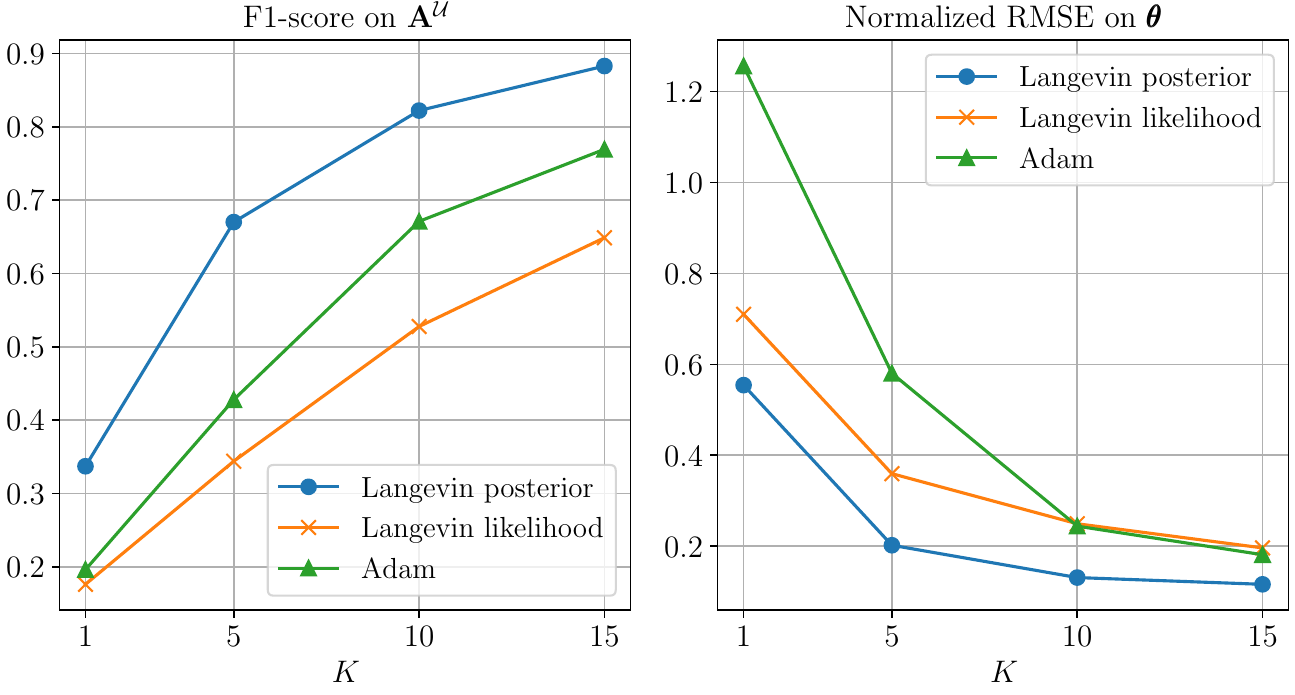}
        \vspace*{-7mm}
        \caption{Estimation performance for grid graphs with a second-order polynomial graph filter.}
        \vspace*{-1mm}
        \label{fig:grids}
    \end{minipage}\hfill
    \begin{minipage}{0.49\textwidth}
        \centering
        \includegraphics[width=\textwidth]{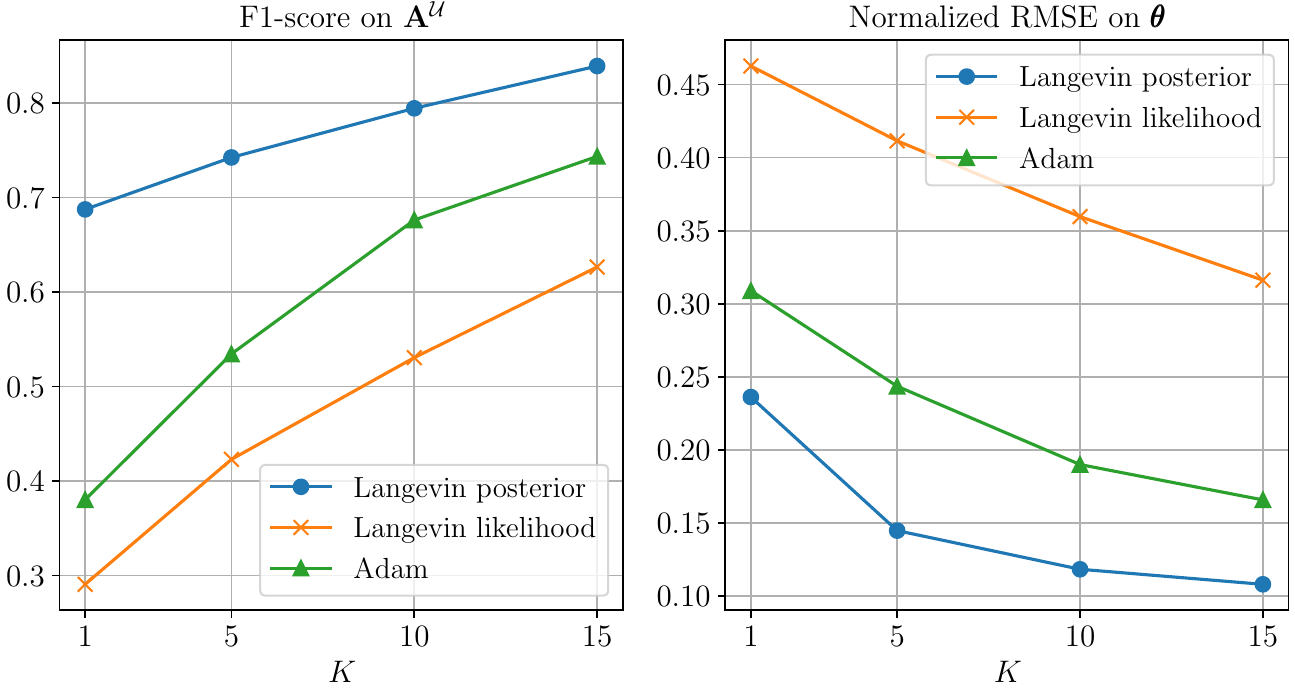}
        \vspace*{-7mm}
        \caption{Estimation performance for ego-nets with a heat diffusion graph filter.}
        \vspace*{-1mm}
        \label{fig:deezer}
    \end{minipage}
\end{figure*}

Under this setting, we propose to jointly estimate $\bbA$ and $\bbtheta$ by running one step of the Langevin dynamics for the former together with one step of a gradient descent algorithm for the latter.
We choose to use Adam for our method.
The final algorithm is described in Algorithm~\ref{alg:annealed_langevin}.
Notice that the score estimator $\bbs(\cdot)$ is an input.
Namely, it is assumed that a GNN has been trained with the desired dataset in order to be able to compute $\bbs\left(\tba, \sigma_l\right) \simeq \nabla \log p\left(\tba\right)$.
After $LT$ steps, the algorithm still yields a continuous matrix $\tbA$.
As we work with unweighted graphs, the algorithm returns $\round{\tbA}$ instead, which stands for an element-wise projection onto $\{0, 1\}$.

\setlength{\textfloatsep}{2pt}
\begin{algorithm}[t]
    \caption{Annealed Langevin for system identification}\label{alg:annealed_langevin}
    \begin{algorithmic}[1]
        \Require $\bbX, \bbY,\bbA^\ccalO, \bbs(\cdot), \{\sigma_l\}_{l=1}^L, T, \epsilon, \tau$
        \State Initialize $\tba$ and $\hbtheta$ at random
        \State $\tbA^\ccalO \leftarrow \bbA^\ccalO$ \Comment{Fix the known values}
        \For{$l \leftarrow 1\; \text{to}\;  L$}
            \State $\alpha_l \leftarrow \epsilon \cdot \sigma_l^2 / \sigma_L^2$
            \For{$t \leftarrow 1\; \text{to}\; T$}
                \State Draw $\bbz \sim \ccalN(\pmb{0}, \bbI)$
                \State Compute $\nabla_\tba \log p(\tba, \bbtheta \mid \bbX, \bbY) $ using~\eqref{eq:score_likelihood}
                \State Compute $\bbs(\tba, \sigma_l)$ using the trained GNN 
                \State $\Delta \leftarrow \nabla_\tba \log p(\tba, \bbtheta \mid \bbX, \bbY) + \bbs(\tba, \sigma_l)$ \label{lst:line:score_update}
                \State $\tba^\ccalU \leftarrow \tba^\ccalU + \alpha_l \Delta^\ccalU + \sqrt{2\alpha_l\tau}\bbz$
                \State Run one Adam step to update $\hbtheta$ with~\eqref{eq:theta_loss} as loss
            \EndFor
        \EndFor
        \\
        \Return $\round{\tbA}$, $\hbtheta$ \Comment{Project each $\tdA_{ij}$ onto $\{0, 1\}$}
    \end{algorithmic}
\end{algorithm}

\section{Numerical results}\label{sec:results}
We run experiments using two different kinds of networks: grid graphs and ego-nets.\footnote{The source code is available at \url{https://github.com/Tenceto/inference_langevin}.}
The first dataset is synthetically generated, while the second one corresponds to real networks of Eastern European users collected from the music streaming service Deezer~\cite{deezer}.
We compare our method with two other approaches:
\begin{itemize}[leftmargin=*]
    \item \textbf{Langevin without prior}. 
    We follow the exact same procedure as in Algorithm~\ref{alg:annealed_langevin} but only using the likelihood. Namely, we set $\bbs (\tba, \sigma_l) = 0$ in line~\ref{lst:line:score_update} at every step.
    \item \textbf{Adam optimizer}. 
    We approximate the maximum likelihood estimator by stochastically optimizing~\eqref{eq:score_likelihood}.
    In each step, we clip the values of $\bbA$ between $0$ and $1$, so that sparsity is enforced.
\end{itemize}

We use $L=10$ noise levels, evenly spaced between $\sigma_1 = 0.5$ and $\sigma_L = 0.03$, and $T=300$ steps per level.
We set the step size at $\epsilon = 10^{-6}$ and the temperature at $\tau=0.5$.
The learning rate of the Adam optimizer is set at $0.01$.
The input signals $\bbx$ are generated as i.i.d. random variables such that $x_i \sim \mathrm{Unif}[-10, 10]$.
The measurement noise has a variance of $\sigma_\varepsilon^2 = 1$.
We report the F1-score of the estimation of $\bba^\ccalU$, and the normalized RMSE for the coefficients of the graph filter.
The reported metrics correspond to averages over $100$ simulations.

\vspace{1mm}
\noindent {\bf Grids.} 
We consider grids of different heights and widths such that $N \in [40, 50]$.
Additionally, in order to add randomness to the graphs, we add between $2$ and $5$ edges randomly between any pairs of nodes.
After generating the graphs, we assume that $25\%$ of the values in $\bba$ are unknown (they belong to $\ccalU$), and we try to estimate them.
We trained the EDP-GNN with $|\ccalA|=5000$ graphs generated with the described procedure to get the estimator $\bbs(\cdot)$.
The filter we use in this case is a second-order polynomial $h_{\bbtheta}(\bbA) = \theta_0 \bbI + \theta_1 \bbA + \theta_2 \bbA^2$, where the coefficients are generated independently as $\theta_i \sim \mathrm{Unif}[-0.1, 0.1]$.
Results are shown in Fig.~\ref{fig:grids}.

\vspace{1mm}
\noindent {\bf Deezer ego-nets.} 
We use the graphs from the original dataset such that $N \leq 25$.
We used $|\ccalA|=2926$ to train the EDP-GNN.
We assume $50\%$ of the values in each $\bba$ to be unknown.
The graph filter used to generate the signals is a heat diffusion $h_{\bbtheta}(\bbA) = \exp(-\theta \bbL)$, where $\bbL = \mathrm{diag}(\bbA \pmb{1}) - \bbA$ is the graph Laplacian.
The filter coefficient is generated as $\theta \sim \mathrm{Unif}[0.3, 0.7]$.
The results of the simulations are shown in Fig.~\ref{fig:deezer}.

\vspace{1mm}
As expected, in both test cases, the estimation performance of all methods improves as $K$ increases.
We observe in all cases that introducing the prior information effectively increases the accuracy in the prediction of both $\bbA^\ccalU$ and $\bbtheta$.


\section{Conclusions}
We introduced an algorithm based on annealed Langevin dynamics for network inference, leveraging prior knowledge about graph distributions. 
This knowledge is incorporated through training a GNN on diverse graphs, regardless of their size or prior distribution.
Our approach provides a sample from the posterior distribution as an estimator, in contrast to the usual point estimation approaches.

Preliminary experiments across various network types revealed that combining observed graph signals with prior knowledge outperforms using only observed data, regardless of the graph filter's functional form.
Current work includes the formulation of theoretical guarantees, further simulations that encompass synthetic and real-world data, and additional comparisons against existing topology inference methods.

\bibliographystyle{IEEEbib_abbrev}
\bibliography{refs}

\end{document}